\documentstyle[aps,twocolumn]{revtex}
\def\dd{\mbox{\small-}}

\begin{document}
\title{Theory of alkali metal adsorption on close-packed metal surfaces}
\author{Catherine Stampfl and Matthias Scheffler}
\address{
Fritz-Haber-Institut der Max-Planck-Gesellschaft, Faradayweg 4-6, D-14\,195
Berlin-Dahlem, Germany
}
\maketitle

\begin{abstract}
Results of recent density functional theory
calculations for alkali metal
adsorbates on close-packed metal surfaces are discussed. Single adatoms on the
(111) surface of Al and Cu are studied with the self-consistent surface
Green-function  method by which the
pure adsorbate-substrate interaction may be analyzed.
Higher coverage ordered adlayers
of K on Al\,(111), Na on Al\,(111), and Na on Al\,(001)
are treated
using the {\em ab-initio} pseudopotential
plane wave  method which affords the prediction of
coverage dependent stable and metastable adsorbate geometries and
phase transitions of the adsorbate layers.
Together, these studies give insight
and understanding  into current key issues in alkali metal adsorption,
namely, the nature of the adsorbate-substrate bond at low coverage
 and the occurrence
of hitherto unanticipated adsorbate geometries, and the
associated electronic properties.

\end{abstract}

\narrowtext
\section{Introduction}

Alkali metal adsorption systems have served as a theoretical
paradigm for understanding the properties of adsorption on
crystalline surfaces
\cite{gurney,lang1,lang2,lang3a,lang3b,lang3,muscat1,muscat3,williams,lang4}.
This is due to the simple electronic structure of the alkali metal atoms
and also because it was
understood that the adatom would not complicate the system of interest
by intermixing with the substrate, but would
 stay on the surface and occupy a high-symmetry,
highly coordinated position.
Early experiments revealed that these ``simple'' adsorbates had a
marked effect on the properties of the surface on which they
were adsorbed \cite{doebereiner}. For example,
in the seminal work of Taylor and Langmuir \cite{taylor} for the Cs/W system,
it was shown that the work function of the clean surface is reduced by several
electron volts  on Cs adsorption and
that at low coverage almost all the Cs adatoms are desorbed as
positive ions.
In the light of such results,
the alkali metal bond to the substrate was
viewed as a spectacular example of ionic bonding akin to that in
alkali halides \cite{naumovets1}.
As a consequence of these distinct and general
properties, alkali metal adsorption has come to play an important role in
technological  applications. For example, the desirable
properties of the considerable enhancement of
the electron emission of cathodes and the improved
activity and selectivity of heterogeneous catalysts
 \cite{ertl,mross} are both achieved
by the addition of alkali metal atoms to surfaces \cite{aruga1}.

A  simple picture of the interaction between alkali metal
atoms and a metal surface,  and of the resulting
chemisorption bond was proposed by Langmuir
\cite{langmuir}. He assumed that the alkali metal atom transfers
completely its valence $s$ electron to the substrate.
In a more rigorous
theoretical description of alkali metal adsorption
Gurney \cite{gurney}, in 1935, proposed a
quantum mechanical picture applicable at low coverages,
where the discrete $s$ level of the free alkali metal atom
broadens (and shifts) becoming partially
emptied as a result of the
interaction with the substrate states as it approaches the surface  (see
Fig.~\ref{fig1}).
As follows from this picture the alkali metal atom is partially positively
charged which induces a negative charge density in the
substrate giving rise to an adsorbate induced
dipole moment which naturally explains the reduction in the work function
of clean surfaces.
The building up of such adsorbate induced dipole moments
also leads  to the
understanding that the dominant interaction between the adsorbates is repulsive
and
that the adatoms should
form a structure with the largest possible
 interatomic distances compatible with
the coverage. In this description it is expected that
with increasing coverage the adsorbate-adsorbate
distance gradually  decreases, and the electrostatic
repulsion between the adatoms increases.
To weaken this repulsion, i.e., to lower the total energy, some fraction
of the valence electrons
flow back from the Fermi level of the metal to the adsorbate.
Thus, a reduction of the adsorbate-induced dipole moment,
i.e., depolarization takes place.
This picture of Langmuir and Gurney
has been held without
question for many years.

The Langmuir-Gurney model was fully supported
by systematic studies of Lang and Williams \cite{lang3a,lang3b,lang3}, who
performed density functional theory calculations
of isolated adatoms bonded to the surface of a simple
metal (``jellium''),
and by Lang's earlier work \cite{lang1,lang2} on the alkali metal
induced work function change.
In these calculations  the ions of the metal substrate were
approximated by a semi-infinite uniform positive background which terminated
abruptly along a plane, otherwise known as the
``jellium'' model.
A jellium model was expected to represent well
the electron density of
a free-electron-like $s$-$p$ bonded metal \cite{jellium}, and indeed these
studies
demonstrated that
many ground-state
properties of metal surfaces and metal-adatom
systems can be described  in a
 physically transparent manner.
As an example, we show in Fig.~\ref{fig2} the change in work function with
increasing coverage for parameters
corresponding to Na on Al\,(111). The curve possesses
a form similar to that often observed experimentally.
This ``usual'' shape of $\Delta \Phi(\Theta)$
is explained as a consequence of the above-mentioned depolarization of the
alkali metal induced  surface dipole
moment induced by continuous reduction of the adsorbate-adsorbate
distance and corresponds to a rapidly
decreasing work function at low coverage, reaching a minimum at
about $\Theta=0.15$, and subsequently rising towards the value of
the pure alkali metal.

The above description represents a sketch of the ``traditional
view'' of alkali-metal adsorption, which, for example,
is underlying the analysis of experimental results in various articles in the
book of Bonzel, Bradshaw and Ertl
\cite{bonzel89}. However, since 1991 it has become clear
through density functional
theory studies \cite{schmalz1,neugebauer2}, as well as through
 several experimental
investigations, that the interactions and reactions of alkali metal atoms
at metal surfaces are more complicated and exciting than hitherto expected.
The calculations,
which take the full atomic structure into account including
relaxation of atoms from their ideal positions as well as surface
reconstructions,
show that the traditional view of alkali metal
adsorption outlined above, is only  part of the
whole picture, and that we must now adopt an updated view.
Phenomena such as the following may occur:
 1) The alkali metal atoms may not necessarily assume  highly
coordinated sites on the surface
\cite{neugebauer2,lindgren,fisher1,adler,shi,davis,huang,carvalho,over,kerkar1,neugebauer1,stampfl1,nielsen1,scragg}.
 2) The alkali metal adatom may kick out  surface
substrate atoms and adsorb substitutionally
\cite{schmalz1,neugebauer2,kerkar1,neugebauer1,stampfl1,nielsen1,scragg,nielsen11,stampfl2,stampfl3,berndt,fasel1}.
Substitutional adsorption has been shown
in three cases to occur as the result of an irreversible phase
transition from an ``on-surface'' site by warming to room temperature,
without change in the periodicity of the surface unit cell or of the
coverage \cite{stampfl1,nielsen1,scragg,nielsen11,stampfl2,stampfl3,berndt}.
 3) The alkali metal atom may switch site on variation of
coverage \cite{over,neugebauer1,gierer,hertel,hertel2,over2},
 and 4) island formation may occur
\cite{naumovets1,neugebauer1,naumovets2,modesti,hohlfeld,aruga2,andersen1}.
5) There may be a strong intermixing of the alkali metal atom with
the substrate surface \cite{andersen2,kerkar2,stampfl4,burchhardt2},
as for example the formation of
a four layer ordered  surface alloy which had been identified  for Na on
Al\,(111) at a coverage $\Theta_{\rm Na} = 0.5$ \cite{stampfl4,burchhardt2}.
The coverage is defined in this paper such that
for $\Theta=1$, the number of adatoms is the same as the
number of atoms in the clean unreconstructed surface layer.
Alkali metal induced surface reconstructions are well known on the
more open surfaces \cite{somorjai,ch14,behm}
as these clean surfaces are close to a structural instability \cite{heine}.
But that significant reconstruction can occur on close-packed
fcc\,(111) and (001) surfaces had not been expected previously.

The associated electronic properties of
these hitherto unexpected surface atomic arrangements,
 perhaps not surprisingly, also exhibit  behavior
deviating from expectations based on early ideas.
For example, experimental measurements of the change in work function
as a function of alkali metal coverage can be quite different
to the ``usual'' form of Fig.~\ref{fig2},
for example, they may not exhibit a minimum at all, or may
have a minimum at a significantly higher coverage.
In the same manner, the density of states
induced by alkali metal adsorbates may not always correspond to that
expected from the model of Gurney.

Contrary to the idea mentioned
above that the traditional picture be kept as representing {\em part}
of a revised view, are recent theoretical
studies which suggested a different picture of alkali metal adsorption be
adopted \cite{ishida11,ishida2}. In this picture
there is practically
no charge transfer from the alkali metal adatom to the substrate but
rather a {\em polarization} of the adsorbate and the adsorbate-substrate
bond is classified as covalent as opposed to the general and widespread
understanding that it is of ionic nature at low coverages.
This interpretation has seemingly been
supported by experimental studies to the extent that
it was concluded that
the traditional picture be entirely discarded \cite{riffe}.
This is a controversial issue but we consider it a largely artificial
conflict based on semantic differences.
For a recent discussion we refer to Ref.~\cite{scheffler,bormet1}, and a short
summary of some of these arguments is given in Section III below.

The purpose of the present paper is to discuss results of
recent {\em ab-initio} calculations for alkali metal adsorbates on close-packed
metal surfaces. These density functional
theory studies provide
an accurate description of the microscopic electronic structure of
the adatom as well as of the atomic geometry.
In Section~II we give a brief outline of the theoretical methods used.
The results of Lang and Williams \cite {lang3b,lang3},
 Bormet {\em et al.} \cite{bormet1}, and Yang {\em et al.}
\cite{yang},
for calculations of  {\em single } adsorbates on jellium, on Al\,(111), and on
Cu\,(111), respectively  are discussed in Section~III.
These calculations afford
an analysis of the nature of the adsorbate-substrate bond
without it being obscured by the influence of other adsorbates.
The results of these studies do not support a dismissal of the traditional view
and the adoption of a covalent picture of alkali metal adsorption when
adsorbed as isolated atoms.
Rather,
the results show that the traditional view is valid and useful but
should only be applied in the limit of very
low coverage and non-activated adsorption (see Section~IV for
the definition of activated and non-activated adsorption).
 The comparison of the results for different substrates enables
one to establish the influence
of the atomic structure and chemical nature  of the substrate
on the calculated quantities.
In Section~IV we discuss recent results obtained using the
self-consistent pseudopotential plane wave method  for
higher coverages of  alkali metal adsorbates on Al.
{}From these calculations ``unusual'' adsorption sites and phase
transitions have been predicted. The calculations also provide insight
and understanding into the reasons why the unexpected adsorbate
arrangements occur.
In subsections IV\,A and  IV\,B we discuss respectively, the
coverage dependence of the adsorbate geometry,  and the adsorbate-adsorbate
and adsorbate-substrate bonding and
the associated electronic properties.
Section~V contains the conclusion.

\section{Theoretical methods}

To provide some understanding of the two different calculation
procedures, namely, the self-consistent surface Green-function
(SSGF) method and the pseudopotential plane wave method,
used to obtain the results discussed in the present paper, we give a short
description of the theoretical basis.
Both methods employ density functional theory (DFT)
with the local density approximation (LDA) for the exchange-correlation
functional (DFT-LDA).
We will first give an outline
of this, then explain each of the calculation procedures separately.
In the equations below Hartree atomic units are used.

\subsection{Density functional theory}

Density functional theory,
as introduced by Hohenberg, Kohn, and Sham \cite{hohenberg,kohn,sham} has
proved
to be a conceptually and practically useful method for studying
the electronic and atomic structure, as well as elastic
and vibrational properties of many-electron poly-atomic systems.
 This theory represents a firm and exact theoretical foundation
 whereby all aspects of the
electronic structure of a system in a
ground state are completely determined by its electron density $n({\bf r})$.
The total energy of the ground state is given by
\begin{displaymath}
E^{\rm tot} [n({\bf r})]
  = T[n({\bf r})]
    + \int V({\bf r})n({\bf r})\,d{\bf r}
\end{displaymath}
\begin{equation}
    + \frac{1}{2} \int \int \frac{n({\bf r})n({\bf r}')}
         {|{\bf r} - {\bf r}'|}\, d{\bf r}\,d{\bf r}'
    + E_{xc}[n({\bf r})] + E_{\rm ion} \quad  ,
\label{eq2_1}
\end{equation}
where $V({\bf r})$ is the ``external potential'' due to the atomic nuclei
(or the ion cores) and $T[n]$ is the kinetic energy functional
for non-interacting electrons having the density $n({\bf r})$. The
functional $E_{xc}[n]$ contains all the quantum mechanical
many-body effects and is called the exchange-correlation energy. Both
$T[n]$ and $E_{xc}[n]$ are not known  explicitly but at least the
kinetic energy can be evaluated exactly when the density is
written as a sum over orthonormal single-particle functions,
\begin{equation}
n({\bf r}) = \sum_{i=1}^{N} |\psi_{i}({\bf r})|^{2} \quad ,
\end{equation}
where $N$ is the number of electrons in the system.
This gives,
\begin{equation}
T[n({\bf r})] = \sum_{i=1}^{N} \langle \psi_{i}| -
\frac{\nabla^{2}}{2}| \psi_{i} \rangle  \quad  .
\end{equation}
The variational property of the total energy then
leads to the equation,
\begin{equation}
  \{ - \frac{\nabla^{2}}{2} + V_{\rm eff}({\bf r})\} \psi_{i}
  = \epsilon_{i}  \psi_{i} \quad ,
\label{eq2_12}
\end{equation}
where
\begin{equation}
 V_{\rm eff}({\bf r}) = V({\bf r}) + V_{H}({\bf r})
 + V_{xc} ({\bf r})
\label{eq2_13}
\end{equation}
which is to be solved self-consistently.
The Hartree and exchange  correlation potentials are,
\begin{displaymath}
V_{H}({\bf r})=\int \frac{n({\bf r}')}{|{\bf r} - {\bf r}'|} \, d{\bf r}'
\end{displaymath}
and
\begin{displaymath}
V_{xc}({\bf r})=\frac{\delta   E_{xc}[n]}{\delta  n({\bf r})} \quad .
\end{displaymath}
The Coulomb energy associated with
interactions among the $M$ nuclei (or ions) at positions ${\bf R}_{I}$
is given by
\begin{equation}
E_{\rm ion} = \frac{1}{2}\sum_{I,J,\,I \neq J}^{M, M} \frac{Z_{I}Z_{J}}{|{\bf
R}_{I} -
{\bf R}_{J}|}  \quad .
\end{equation}
The only unknown quantity is the exchange-correlation
energy functional, $E_{xc}[n]$, and the quality of the solution of the
full many-body problem is only limited by the quality of the approximation used
for it.
The exchange-correlation energy can be written as,
\begin{equation}
E_{xc}[n({\bf r})]= \int \epsilon_{xc}[ n({\bf r})] \, n({\bf r})
\, d{\bf r} \quad .
\end{equation}
A simple yet effective way to evaluate
$E_{xc}[n]$ is to use the
LDA \cite{kohn} which means that the
exchange-correlation energy density functional,
$\epsilon_{xc}[n]$,
  is approximated by the corresponding
expression for the homogeneous electron gas.
That is,
\begin{equation}
\epsilon_{xc}[n] \equiv \epsilon_{xc}^{\rm LDA}(n)
\end{equation}
where $\epsilon_{xc}^{\rm LDA}(n)$
 is the exchange-correlation energy per particle of the many-body
system which is known most accurately, namely, the uniform
electron gas. Experience from numerous calculations show that for well bonded
situations the LDA represents a good description of the
quantum-mechanical many-body interactions in poly-atomic systems,
and calculated electron densities, $n({\bf r})$, atomic geometries, and total
energy differences are in fact very reliable. In actual calculations,
 often some additional, more severe
{\em numerical} approximations are applied. The latter can (and should) be
carefully
tested.

\subsection{The self-consistent surface Green-function method}

The self-consistent Green-function
(SSGF) method is outlined briefly here and is presented
in detail in Refs.~\cite{scheffler,bormet1,bormet2}.
It gives an accurate and efficient description for {\em isolated}
 adsorbates on  {\em semi-infinite}
substrates.
Using this method the total and adsorbate induced
 density of states (DOS), charge densities and density
differences can be obtained, as well as the total energy,  total energy
change, and forces on the adsorbate.
In this approach the system is split into a {\em reference system} and
a perturbative part with a perturbation potential,
 $\Delta \,V_{\rm eff}[n]$, which has to be evaluated self-consistently.
The reference system is taken to be the ideal, clean surface and the
corresponding Green-function, $G^{0}$, is calculated employing
the layer-Korringa-Kohn-Rostoker (KKR) Green-function
method \cite{kambe}. The full Green-function of the adsorbate
system is given by the Dyson equation,
\begin{equation}
G = G^{0} + G^{0}\, \Delta V_{\rm eff} \, G \quad .
\label{eq2_2}
\end{equation}
Subtracting $G^{0}$ from both sides of equ.~\ref{eq2_2} gives,
\begin{equation}
\Delta G = G - G^{0} = G^{0}\, \Delta V_{\rm eff} \, G
\end{equation}
\begin{equation}
= G^{0}\, \Delta V (1 - G^{0} \, \Delta V_{\rm eff} )^{-1} G^{0} \quad .
\label{eq2_3}
\end{equation}
The perturbation potential, $\Delta V_{\rm eff}$,
is written as,
\begin{equation}
\Delta V_{\rm eff} [n({\bf r})] = V_{\rm eff}[n({\bf r})] - V^{0} \quad ,
\end{equation}
where $V^{0}$ is the potential defining the
reference system and the Green function $G^{0}$, and
$n({\bf r})$ is the density of the valence electrons.
Thus, $V_{\rm eff}[n({\bf r})]$ is
 the functional of the effective potential as described by the Kohn-Sham
equation (equs. \ref{eq2_12} and \ref{eq2_13}).
The perturbation potential, $\Delta V_{\rm eff}$,
is strongly localized in real space.
This localization allows equ.~\ref{eq2_3}
 to be solved efficiently in a rather small, localized basis set.
In the work discussed below,
 $s,p,$ and $d$ gaussians and three different decay constants
per angular momentum number are used which are centered
on the adatom and on the three nearest-neighbor substrate
atoms \cite{bormet1,yang}.

Once the change in the valence electron density is calculated
self-consistently, the total energy, $E^{\rm tot}$,
as well as the forces, $F=\frac{dE^{\rm tot}}{d {\bf R}}$,
on the adsorbate as a function of its position ${\bf R}$ can
be evaluated \cite{bormet1}.

\subsection{The {\em ab-initio} pseudopotential plane wave method}

A very successful way to solve the Kohn-Sham single particle equation and
to perform accurate calculations for a periodic system is
 achieved when ion cores are treated by {\em ab-initio} norm-conserving
pseudopotentials and a plane waves basis set is employed.
The great advantage of this approach is that the basis set is
{\em unbiased}, i.e., the numerical accuracy can be tested
systematically. Furthermore,
it provides accurate forces on the atoms without complications typically
encountered when other basis functions are used.
 The wave functions are written as,
\begin{equation}
\psi_{j}({\bf k},{\bf r}) =  \sum_{|{\bf G} + {\bf k}|
 \leq G_{\rm max}}
c_{j,{\bf G} + {\bf k}} {\rm exp}(i({\bf G } + {\bf k}) \cdot {\bf r})
\quad  ,
\end{equation}
where ${\bf G}$ are the reciprocal
 lattice vectors.
Because the coefficients,
$c_{j,{\bf G} + {\bf k}}$, for plane waves with high kinetic
energy decrease and become negligible,
the plane wave basis set can be truncated to include only
plane waves that have kinetic energies less than some particular cutoff
energy, $\frac{\hbar^{2}}{2m}|G_{\rm max}|^{2} = E_{\rm cut}$.
Thus, the cutoff energy defines the quality of the basis set and it is one
of the important numerical parameters of the calculations for which it must be
carefully tested that they do not affect the results.
In the pseudopotential plane wave method a crystal surface is modeled by a
large
supercell which contains a slab a certain
 number of atomic layers thick which is
surrounded
by vacuum. This supercell is repeated throughout space. The surface
is then periodic in the plane of the surface on the scale of the supercell,
but the periodicity perpendicular to the surface,
imposed by the supercell, has no
influence because for a sufficiently thick vacuum region
the different slabs will not interact.
However, care must be taken that the slab is thick
enough so that its two surfaces are practically decoupled from each other.
For a further description of the numerical details
we refer to Refs.~\cite{neugebauer2,stumpf2}.

\section{Isolated adsorbates}

By studying {\em isolated} adatoms on surfaces
the nature of the pure adsorbate-substrate bond can be analyzed.
Self-consistent DFT-LDA calculations for isolated adatoms on a
semi-infinite jellium surface have been performed by
Lang and Williams \cite{lang3b,lang3}.
Bormet {\em et al.} \cite{bormet1} carried out an analogous study
for a representative set of isolated adsorbates
on Al\,(111). In the latter work, instead
of using a jellium model for the Al substrate, the atomic structure of
the substrate is taken fully into account.
Therefore, any significant differences in the results of the two studies is
directly attributable to the treatment of the substrate.
As a complementary study to the adsorbate on Al\,(111)
 systems, Yang {\em et al.} \cite{yang} investigated isolated
adsorbates on Cu\,(111).
{}From comparison with the results for Al\,(111), this work allows
the influence of the Cu $3d$ electrons on the adsorbate bonding
to be identified.
 In both of these latter calculations the self-consistent
surface Green-function (SSGF) method was used.
The set of adsorbates from the
third row of the periodic table, were
chosen in order to establish chemical trends. We concentrate here, in
particular,
on the case of alkali metal adsorption, but briefly describe the
results for the
other adsorbates, namely,  Si and  Cl in order to highlight the very
different and characteristic behavior of the
adsorbates which
is found to be highly correlated with
the  electronegativity
of the adsorbate with respect to the substrate.

\subsection{Nature of the bond
in the limit of very low coverage}

For adsorption on Al\,(111) and Cu\,(111) the single
 adsorbates were placed in the fcc-hollow site, and
in order to compare with the results of Lang and Williams \cite{lang3b,lang3},
the surface was taken to be unrelaxed.
This is, in fact, a good approximation because for non-activated adsorption
the actual substrate relaxation is rather small.
The vertical height of the adsorbates was optimized
by searching for the minimum of the total energy.
{}From the study of Ref.~\cite{bormet1}
it was found that the equilibrium heights of the adsorbates above
the surface are smaller
for the Al\,(111) substrate  than for the jellium  substrate.
With respect to the center of the top aluminum layer the heights
are approximately 3, 20, and 20\% smaller on Al\,(111)
than on jellium for Na, Si, and Cl, respectively.
For jellium the ``center of the top layer'' is
defined to lie half a bulk inter layer spacing of Al\,(111) below
the jellium edge.
The larger values obtained in the jellium calculations are due
to neglecting the atomic structure of the
substrate which has the ability to bind the adsorbates more strongly.
It is interesting that for Si on a jellium
surface and using first-order perturbation theory for the crystal lattice,
an equilibrium height is obtained which is identical to the full
SSGF result \cite{lang3,bormet1}.
On comparison of the results for adsorption on the Cu and Al substrates
it is found that the jellium$\rightarrow$aluminum trend continues:
the effective radius of the studied adsorbates
on Cu\,(111) is smaller than on Al\,(111) by  approximately
11, 15, and 17\%, respectively, for Na,  Si,  and Cl.
Clearly, in this case it is more meaningful to compare effective
radii than inter layer spacings due to the different lattice constants
of Al and Cu.
The smaller values reflect
 the existence of a stronger bonding
of the adsorbates to the substrate for Cu\,(111) as compared to
Al\,(111). This is identified as being due to the Cu
$d$ electrons \cite{yang}.

Adsorbate induced density of states (DOS) and charge density difference
plots are informative
in discerning the nature of the adsorbate-substrate bond.
The adsorbate induced DOS
indicates whether the state of interest is occupied,
unoccupied, or partially occupied, while
charge density difference plots show whether there is a shift of
charge density from the adsorbate towards the substrate, vise versa, or a
sharing of charge between the adsorbate and the substrate.

The adsorbate induced DOS
obtained from the SSGF calculations of Bormet {\em et al.}
\cite{bormet1}
and those of Lang and Williams \cite{lang3}
for adsorbates on jellium, corresponding to
a high electron density metallic substrate ($r_{\rm s}=2$ bohr),
 are shown in Figs.~\ref{fig3}a and \ref{fig3}b, respectively.
The adsorbates investigated
in the work of Bormet {\em et al.} \cite{bormet1}
were Na, Si, and Cl
and in that of  Lang and Williams \cite{lang3}
they were Li, Si, and Cl.
The results of both of these calculations agree qualitatively and
 may be interpreted as follows: For both Na  and
 Li  the adsorbate resonance lies well above
the Fermi level and is thus largely unoccupied.
This indicates that the valence electron of the alkali metal atom
 (or part thereof) has been
transferred to the substrate and the adatom is partially
positively charged.
In an opposite manner, on adsorption of
chlorine, the resonance in the curve corresponding
to the Cl $3p$ resonance lies well below the Fermi level.
Thus, this resonance is fully occupied which implies that
a transfer of electronic  charge density from the substrate to the Cl
adatom has taken place; the adsorbed Cl atom
is partially negatively charged.
For the adsorption of an isolated Si atom it can be seen from the jellium
calculations that the Si $3p$ resonance lies just at the Fermi level,
which implies that it is about half occupied.
Lang and Williams \cite{lang3} have shown that the states of the energetically
lower half of the $p$-resonance are bonding between the adatom
and the substrate and that the energetically higher states
are anti-bonding. Because the Fermi level cuts the $p$-resonance approximately
at its maximum, the bonding nature of Si is covalent.
The results for Si on Al\,(111)
show also that the bond is covalent.  In this
case, however, more structure occurs in the DOS than in the
jellium calculations. This arises because  the atomic structure
of the substrate leads to band structure
effects (clearly reflected by the structure of the bulk DOS at
$E_{F}$ in Fig.~3a),
 and these induce a splitting of the bonding and anti-bonding
states,  and the adatom density of states
exhibits a minimum at the Fermi level \cite{bormet1}.
Similarly to the jellium substrate, the
 Fermi level cuts the Si $3p$ induced DOS roughly in the middle.

The behavior identified in the adsorbate induced
DOS triggers the distribution of the
valence electrons.
In Fig.~\ref{fig4} we show
the valence electron density for the chemisorption of Na, Si,
and Cl on Al\,(111).
 The charge transfer from the Na atom to the
substrate,
 as reflected by the adsorbate induced DOS, is clearly visible in
 Fig.~\ref{fig4} where
it can be seen that
there is an increase of the electron density between the Na adatoms and
the topmost layer of the  Al substrate, and that on the
vacuum side of the Na atom the
electron density has almost disappeared. For Si, as expected from
the adsorbate induced DOS,
a directional covalent bond is present between the Si adatom
and the nearest-neighbor substrate atom. Furthermore,
it can be seen that
the maximum of the charge density of the chemisorption bond is
closer to the more electronegative Si atom.
In the case of Cl on Al the charge density distribution around
the adsorbate is almost spherical.

In Fig.~\ref{fig5}  we consider
density difference plots.
These plots display the
electron density of the adsorbate system
minus the density of the  clean substrate and minus the electron
density of the free adatom.
The results for the Al and Cu substrates show a number of
similar features, and some interesting differences.
Firstly, it can be noted that in each case the perturbation to the system
caused by the adsorbates does not reach very far into
the metal substrate.
 The interior is essentially identical to that
of the clean surface for layers deeper than the second.
We emphasize that this localization
holds for the electron density {\em perturbation} but
not for individual wave functions.
For Na adsorption, perhaps the most noticeable feature is
 the charge cloud between the adsorbate and the
substrate which
reaffirms that charge is displaced from the vacuum side of the adsorbate
towards the substrate side.
These results are consistent with a
partially positively charged adatom which sits on
a negatively charged site. This constitutes a surface
dipole which locally decreases the work function.
The opposite situation is found for Cl,  which is negatively charged and
sits on an adsorption site that is positively charged.
In this case charge has moved from the substrate towards the Cl atoms and
 the local work function therefore will be increased relative to the value
of the clean surface.
For both substrates Si appears covalently bound with a slight electron transfer
towards the adatom.
Comparing in more detail the results for Al and Cu, we see that there is
more charge between the Na adatom and the top  layer of the
Cu substrate. Also, there is a greater depletion of
charge at the vacuum side. These effects are consistent with
the fact that the electronegativity difference between Na and Cu
 $(0.93 - 1.90 = -0.97)$ is larger than that between Na and
Al $(0.93 - 1.61 = -0.68)$.
The Si atom, which clearly forms a covalent bond with both substrates,
is slightly more electronegative than Al (by 0.27).
 In this respect, it can be seen that
more charge resides on the Si atom when on Al than when on Cu for which
the electronegativity difference is zero.
The result for Cl on Cu displays some features similar
Si  on Cu, namely, the pile up of charge density between the Cl adatom
and the nearest Cu atom. This structure is practically absent for Cl  on Al.
In all cases for the adsorbates studied,
the adsorption on Cu exhibits some structure in the
valence electron density change
near the nucleus of the Cu atom closest to the adsorbate. This reflects the
participation of the Cu $d$ electrons in the bonding.

To summarize,
Green-function calculations for single adatoms on
jellium, Al\,(111), and  Cu\,(111) substrates show qualitatively the same trend
for the different adsorbates and this is well described
by the adatom electronegativities compared to the electronegativity
of the substrate
atoms.
For single alkali metal atoms on these surfaces,  all the calculations
support the charge transfer picture
emphasized in the model of Gurney as
providing a useful and appropriate description of
the situation for alkali metal adsorption at
very low coverage.
Admittedly (see also Refs. \cite{scheffler,bormet1})
the picture is useful and appropriate, but it is not
unique. A (partially) charged entity in front of a metal
surface will necessarily induce a screening charge density at the metal
surface  --  the
quantum mechanical realization of the image charge. This screening charge
density is located between the adatom and the substrate surface.
 Therefore, from a
purely mathematical point of view, the adatom induced electron density can
be described in three different ways: $i)$ in terms of orbitals of the
adatom, $ii)$ in terms of orbitals of the substrate, and $iii)$ by taking both
orbital sets into account. This ambiguity is due to the fact that the sum
of the two Hilbert spaces of the adatom and of the substrate is over-complete.
Whereas a mathematical analysis of the adsorbate induced electron density
leaves this freedom, the chemical interpretation of the DOS, as well as its
effect on the charge density, as discussed above, clearly support the
qualitative picture of Gurney. We note, however, that this
picture applies only to low coverages and an unreconstructed
substrate. Otherwise, as
will be seen in Section~IV,
some significant modifications take place.

\section{alkali metal adlayers on Al}

One of the interesting characteristics of alkali metal adsorption is the
drastic change of several physical quantities as a function of coverage.
The most obvious of these properties are perhaps the surface
atomic geometry and the work function. With respect to the former,
different commensurate phases are formed by  alkali metal adsorbates on
metal surfaces. Some of these structures can be understood
as being due to a repulsive lateral adsorbate-adsorbate
interaction, causing the adatoms to sequentially
arrange into commensurate
phases of ever growing compactness  \cite{naumovets1}.
In addition, as mentioned in the introduction, behavior quite different to
these arrangements has been reported, as well as
a new class of unexpected
geometries becoming evident, namely, those of
{\em substitutional} and {\em surface alloy} structures.

To learn about the strong coverage dependence of the
surface atomic arrangements
and related properties, calculations for
higher coverages of alkali metal adsorbates
on metal surfaces have been performed
for
K on Al\,(111) \cite{neugebauer2,neugebauer1,neugebauer3},
Na on Al\,(111) \cite{neugebauer2,neugebauer1,stampfl4,neugebauer3},
 and Na on
Al\,(001) \cite{stampfl2,stampfl3}.
The calculations show that on
increasing the coverage from  very low coverage,
where the alkali metal adatoms
are uniformly spread over  the surface,
to higher coverages,  the adsorbate-adsorbate
interactions weaken the adsorbate-substrate bond
because additional bonds are built up within the
adlayer. We note that this is a
general feature of
bonding in poly-atomic systems: When the valence electrons are shared among
many bonds, each
single bond becomes weaker, and an increase of the bond length
results. In the case of alkali metal adlayers this implies that
a formation of lateral bonds in the adlayer will weaken the
adsorbate-substrate bonds and increase the adsorbate-substrate
bond length. This interplay/competition between adsorbate-adsorbate
and adsorbate-substrate bond formation
can give rise to interesting adsorbate arrangements and structural
phase transitions.
The results for the three systems are compared which
affords changes due to different
alkali metal atoms on the same surface to be identified as well
as changes due to different substrate surfaces for the same adsorbate.

It is necessary to distinguish ``non-activated''
and ``activated'' adsorption.
 The former corresponds to adsorption
on an  only relaxed surface (small rearrangements  of substrate atoms)
 where no energy barriers need
to be overcome to reach the adsorption geometry.
The latter refers
 to adsorption involving activated processes
where it is likely that an energy barrier
 must be overcome in order to reach the adsorption geometry, and the
substrate surface undergoes a significant
 reconstruction.
In the present work,
non-activated and activated adsorption may be regarded as corresponding
to structures formed at low temperature and
at  room (or higher) temperature, respectively.

\subsection{ Adsorbate geometry}

By calculating the adsorption energy for various
coverages and adsorption sites, detailed insight about stable and metastable
adsorbate geometries and structural phase transitions
can be obtained. Furthermore,
an accurate knowledge of the adsorbate and substrate atomic geometry
is crucial for any additional
analysis of the adsorbate properties, as for example, the surface electron
density of states, chemical reactivity,  and adsorbate induced work function
changes. The technical details of the  density functional theory calculations
are given elsewhere \cite{neugebauer2,stumpf2}.

The stable and metastable geometries are determined by an automated
movement of the atoms of the adsorbate and substrate surface on the
Born-Oppenheimer energy surface into the total energy minimum.
For the analysis,
it is necessary to define the {\em adsorption energy
per adatom}. We will write this definition down using indices which refer
to the adsorption of Na on Al\,(111),
but a
transformation to other systems is obvious.
The adsorption energy is the difference
of the total energy of the adsorbate system and the total energy
of the clean, unreconstructed Al\,(111) substrate together with a free,
neutral Na atom.
Because total energies of atoms have negative
values, the present and usually applied definition,
associates exothermal adsorption with a positive adsorption
energy \cite{sign}.
The adsorption energy for the {\em on-surface}
(non-activated) adsorption is given by
\begin{equation}
E_{\rm ad}^{\rm Na/Al\,(111)}  = E^{\rm Al\,(111)}
   + E^{\rm Na \dd atom} - E^{\rm Na/Al\,(111)} \quad ,
\label{eq3_1}
\end{equation}
where $E^{\rm Na/Al\,(111)}$ is the total energy per adatom of the adsorbate
system,
$E^{\rm Al\,(111)}$ is the total energy of the clean Al\,(111) substrate,
and $E^{\rm Na \dd atom}$ is the total energy of a free Na atom.

For substitutional adsorption the definition is in principle
identical, however, one has to take into account that the
total energy $E^{\rm Na/Al\,(111)}$ now corresponds to a system with
a substitutional Na atom for which the kicked off surface Al atom
has diffused to a kink site at a step where it is rebound. Thus, for the
substitutional adsorption the adsorption energy is
\begin{displaymath}
E_{\rm ad}^{\rm Na/Al\,(111) \dd sub}
 = E^{\rm Al\,(111)} + E^{\rm Na \dd atom} - E^{\rm Na/Al\,(111) \dd sub}
\end{displaymath}
\begin{equation}
 - (E^{\rm Al \dd atom} - E_{\rm coh}) \quad .
\label{eq3_2}
\end{equation}
Here we have written the total energy of the displaced Al atom
which is rebound at a kink site, as the sum of the
total energy of the
free Al atom and the negative of the cohesive energy . Thus, we assume
thermal equilibrium and
take into account that this is transmitted by kink sites at steps.
The quantity $E^{\rm Na/Al\,(111) \dd sub}$ in equ. \ref{eq3_2} is the total
energy of the slab with the
adatoms adsorbed in substitutional sites.

It is sometimes useful for a more transparent discussion to describe
the substitutional adsorption in terms of two independent processes,
namely, at first the creation of the surface
vacancy structure, and then the chemisorption of the adatoms  on
the substrate surface with vacancies.
Firstly, the energy required to create the surface
vacancy structure is equal to
the total energy of the (reconstructed) substrate (with
surface vacancies), $E^{\rm vac}$, plus the total
energy of an atom at the kink site  minus the total
energy of the clean and unreconstructed surface,
\begin{equation}
E_{f}^{\rm vac}  = E^{\rm vac} + (E^{\rm Al-atom}
 - E_{\rm coh}) - E^{\rm Al\,(111)} \quad .
\label{eq3_3}
\end{equation}
The vacancy formation energy has typically a positive
value, as it usually costs energy to create a vacancy.
In the second process, the vacancies are occupied
by the adatoms. This allows us to define the
{\em binding energy} per substitutionally adsorbed
adatom, $E_{b}^{\rm sub}$,
\begin{equation}
E_{b}^{\rm sub} = E_{\rm ad}^{\rm Na/Al\,(111) \dd sub} +
E_{f}^{\rm vac} \quad .
\label{eq3_4}
\end{equation}
For the on-surface adsorption the {\em binding} and {\em adsorption}
energies are the same.

In general, when a surface structure is considered which has
a different number of substrate atoms to the clean unreconstructed surface
the required or surplus  substrate atoms are taken from or to kink sites
at steps, using again the fact that in thermal equilibrium
kink sites represent the thermodynamic reservoir for substrate atoms
and establish that the chemical potential equals the cohesive energy.

\subsubsection{Atomic geometries of K on Al\,(111) for $0 < \Theta_{\rm K}
\protect\leq 0.33$}

This subsection describes various adsorbate geometries which are assumed by
potassium atoms on the (111) surface of aluminum, depending on the adsorption
temperature and the adsorbate coverage. At first,
non-activated adsorption is discussed, for which the calculations predict
that at low coverage the
adatoms form a disordered homogeneous
adlayer  with the potassium atoms occupying preferentially on-surface hollow
sites. At $\Theta_{\rm K} \approx 0.15$
a structural phase transition occurs from this homogeneous adlayer
into close-packed islands in which the adatoms occupy on-top positions.
If thermal energy is available to overcome energy barriers,
the adatoms in these islands undergo a site change into a substitutional
geometry. Surprisingly, the surface periodicity is unchanged in this
major disruption of the surface, in which one third of the surface
Al atoms are removed. The subsection is concluded with a brief comparison
of the DFT-LDA results with  recent experimental studies.

The calculated adsorption energy for K on Al\,(111)
for several conceivable adsorption sites and for various coverages
are collected in
Figs.~\ref{fig6}a and \ref{fig6}b, where the latter also gives
the binding energies for substitutional adsorption, $E^{\rm sub}_{b}$, and the
vacancy formation energies, $E^{\rm vac}_{f}$.
These results show that at low coverages the
on-surface threefold hollow position is energetically
preferred.  The rapid decrease of the adsorption energy with
increasing coverage reflects a strong adsorbate-adsorbate repulsion.  This is
consistent with the strong adsorbate induced
dipole moments which repel each other.
The K atoms are expected therefore to be homogeneously
distributed over the surface at low coverages.
For higher coverages the results indicate that a phase transition occurs
to adsorbate islands with a condensed structure of $(\sqrt{3} \times
\sqrt{3})R30^{\circ}$
 periodicity. The local coverage in these islands is $\Theta_{\rm K}=1/3$.
Interestingly,
the adsorption energies for this periodicity
are more favorable
than those with a $(2 \times 2)$ periodicity and coverage
$\Theta_{\rm K}=1/4$.
It can be therefore concluded that for coverages
smaller than $\Theta_{\rm K} \approx 0.15$ the Gurney picture and
the dipole-dipole repulsion (and depolarization) appear to describe the physics
of the adlayer appropriately. However, when the coverage is increased,
there is a crossover point,
where a different mechanism becomes energetically
preferable: Instead of keeping the largely ionic adsorbate-substrate
bonds and the energetically unfavorable
repulsive interaction between adatoms, it becomes energetically
beneficial to build up a close-packed potassium layer with a metallic bonding.
Together with this, a reduction of the strength of the
adsorbate-substrate bonding takes place.
We had noted above that this is a very general
phenomenon (roughly the bond strength scales with the square root of
the coordination number). Thus, with the increase of
coordination, the adsorbate-substrate bond becomes weaker and longer.
The calculations show that in this condensed structure
with ($\sqrt{3} \times \sqrt{3}$) periodicity, the on-top position is now
energetically slightly more favorable than the on-surface hollow position.
Thus, for
non-activated adsorption the theory predicts a structural phase transition from
a homogenous, rather open adlayer with adatoms preferentially occupying
the on-surface hollow sites, into two-dimensional islands with a ($\sqrt{3}
\times
\sqrt{3}$) structure
where the adatoms occupy on-top sites.
 This phase transition
is  indicated by the dashed line in Fig.~\ref{fig6}a.

Figure \ref{fig6}a shows that the adsorption energies of all the geometries
considered with the $(\sqrt{3} \times \sqrt{3})R30^{\circ}$
periodicity are very similar, which indicates that
at room temperature  substitutional adsorption becomes energetically possible.
In order to reach the substitutional geometry an energy barrier must
be overcome to reconstruct the substrate surface. We will
at first analyze the on-surface $(\sqrt{3} \times \sqrt{3})R30^{\circ}$
adlayer and then we come back to the
substitutional adsorption.

For low temperature (i.e.,  non-activated) adsorption,
Fig.~\ref{fig6}a predicts that simultaneously with the island formation
a site change occurs;  the adatoms switch from on-surface
hollow sites to on-top sites, which also
involves a noticeable relaxation of the substrate \cite{neugebauer2,stampfl1}.
In particular, the Al atoms in the first substrate layer
beneath the on-top K atoms are moved down and
the Al atoms,  between the on-top K adatoms
are shifted upwards, relative to the geometry of the unrelaxed
substrate surface (see Fig.~\ref{fig7}a). It is important to note that the
atomic radius of potassium is
significantly larger than that of Al. This suggests, what is
fully supported by the calculations, that K adatoms experience a rather small
corrugation of the Al\,(111) surface. In fact, with the above-mentioned
substrate relaxation, and because of the smaller K-Al bond length in the
on-top geometry (due to the lower coordination of the K adatom)
potassium  at on-top sites come even closer to the surface than for the
on-surface hollow sites.
As a consequence, a better embedding of the adatoms  in the substrate electron
density
is achieved.

In the substitutional geometry, which can be reached when thermal energy
enables energy barriers to be overcome, it is found that
because of the larger size of the  alkali metal atoms
compared to the Al atoms,
the K adatoms jut out of the surface
(see Fig.~\ref{fig7}b).
The effective  radius of K in the substitutional site
is larger than it is in the on-top site as can
be seen clearly from comparing
Figs.~\ref{fig7}a and \ref{fig7}b.
This trend reflects again the relation between the bond strength and
the local coordination: the higher the coordination, the weaker
each bond is, and the longer each bond length is.
The DFT-LDA results \cite{neugebauer2} gives this increase in bond length as
about 16\%, and a LEED analysis \cite{stampfl1} gives
19\%.
With respect to the interatomic distances of the on-top
geometry, the nearest-neighbor K-Al distances are 3.38 and 3.23~\AA, and
for the substitutional site the values
are 3.70 and  3.58~\AA, as obtained from the DFT-LDA calculations
and the LEED  analysis, respectively.
A rough theoretical
estimate  of the temperature at which the transition from on-top
to substitutional adsorption could occur with a measurable rate
was obtained \cite{stampfl1} by assuming that the transition state of the
reaction path is such that the K atoms
are in substitutional sites and the ejected Al atoms are  in isolated
positions on the surface (not yet rebounded at steps).
This transition state corresponds to an energy barrier of $E^{b}=0.8$~eV.
The Arrhenius equation, with a reasonable (but rough)
estimate of the pre-exponential factor for the kick out
event, then yields a transition temperature of 220~K.

The nature of the adsorbate-adsorbate interaction
for the substitutional site
and the mechanism actuating the
island formation with a  substrate reconstruction
with the $(\sqrt{3} \times \sqrt{3})R30^{\circ}$ structure
can be learnt from Fig.~\ref{fig6}b.  It shows
that the vacancy formation energy is particularly low for the
$(\sqrt{3} \times \sqrt{3})R30^{\circ}$ surface vacancy structure compared
to lower concentration vacancy structures.
This behavior reflects an attractive vacancy-vacancy interaction,
which is actuated by the fact that the group III element Al can easily
assume covalent bonds. For the $(\sqrt{3} \times \sqrt{3})R30^{\circ}$ surface
vacancy structure, which is equivalent to that of a graphite layer,
the coordination of the surface Al adatoms
indeed gives rise to the formation of directional bonds between the
surface Al atoms.
Figure~\ref{fig6}b shows that the binding energy of K in a surface
vacancy is relatively
insensitive to the coverage and even exhibits
a slight repulsive interaction between the K adatoms as indicated
by the small decrease in binding energy of the
$(\sqrt{3} \times \sqrt{3})R30^{\circ}$
structure as compared to the lower coverage structures.
Thus it is the relative stability of the $(\sqrt{3} \times
\sqrt{3})R30^{\circ}$ surface vacancy structure,
which is the driving force for the
island formation with a substitutional
$(\sqrt{3} \times \sqrt{3})R30^{\circ}$
structure \cite{neugebauer2}.

Turning now to experimental studies
we like to mention the low energy electron diffraction (LEED)
study \cite{stampfl1}, of which some results were already noted above.
This investigation clearly proved that the $(\sqrt{3} \times
\sqrt{3})R30^{\circ}$ structure
formed at low temperature contains K atoms in the on-top site, which,
on warming to room temperature transforms irreversibly into a
$(\sqrt{3} \times \sqrt{3})R30^{\circ}$  structure where
the K atoms adsorb substitutionally.
That is, even if the temperature is lowered the K atoms
will remain in substitutional sites.

The two different phase transitions predicted by the theory (on-surface hollow
sites $\rightarrow$ on-top site island formation, and on-surface $\rightarrow$
substitutional geometry) are also consistent with
high resolution core level spectroscopy (HRCLS) measurements
\cite{andersen1}  which show that island formation starts
at $\Theta_{\rm K}$ = 0.12 and 0.10 at low and room temperature, respectively.
Similarly, a recent experimental study of the temperature dependence of
optical second-harmonic generation \cite{wang}
found that at coverage $\Theta_{\rm K}$ =0.06 no
phase transition occurred when warming to room temperature whereas at
$\Theta_{\rm K}$
= 0.12 the transition took place. These experiments thus confirm the
theoretical estimate, which was $\Theta_{\rm K} \approx 0.15$, quite well.
The temperature at which the phase transition into the
substitutional adsorbate islands takes place was measured as 210~K
\cite{wang} which is in surprising agreement with the rough
theoretical estimate \cite{stampfl1} noted  above.

\subsubsection{Atomic geometries of Na on Al\,(111) for $0 < \Theta_{\rm Na}
\protect\leq 0.56$}

For Na on Al\,(111)
the same adsorbate geometries and coverages
have been investigated by DFT-LDA calculations as for K on Al\,(111), with the
addition  of
a  structure with a  $(4 \times 4)$ periodicity corresponding to
coverage $\Theta_{\rm Na}=9/16$ which represents a densely packed hexagonal
adlayer
with nine Na adatoms per surface unit cell all occupying different (mostly
low symmetry)  on-surface sites. Furthermore, a variety of many different
structures at $\Theta_{\rm Na}=0.5$ was analyzed, because experiments
had indicated a strongly intermixed surface geometry, without, however, giving
a clue about the structure.
The calculations predict phase transitions which are very similar to those of
K/Al\,(111), but some aspects are
noticeably different, due to the fact that Na is  smaller than potassium
although still  larger than Al. The $\Theta_{\rm Na}=0.5$
structure represents the most significant difference to the potassium
adsorbate, but it is also unique with respect
to any other adsorbate system: This structure may be described as
an ordered four-layer surface alloy on an unreconstructed
Al\,(111) substrate.

We start with a discussion of the theoretical results for the
non-activated adsorption; they are summarized in Fig.~\ref{fig8}a
and can be described as follows:
For low coverages the
on-surface hollow site  is  energetically most
favorable and, in a manner similar to that for K on Al\,(111),
 a strong repulsive adsorbate-adsorbate
interaction is obvious.
In this case,  as opposed to K on Al\,(111) the on-top site
is {\em significantly} less favorable than the on-surface hollow
 site at all coverages considered.   This can be attributed to the
smaller
size of the Na atoms which experience a more highly corrugated
potential energy surface due to the substrate atomic structure
 than the larger K atoms.  Figure~\ref{fig8}a shows that
the condensed structure  (the atomic geometry is displayed
in Fig.~9a) for $\Theta_{\rm Na}=9/16$ is energetically more
favorable than homogeneous adlayers of Na
for coverages larger
than approximately $\Theta_{\rm Na}=0.1$.
Accordingly, these results imply that
for very low coverages
the adsorbates occupy on-surface hollow sites and are uniformly distributed
over the surface (homogeneous adlayers)
but for coverages greater than about $\Theta_{\rm Na}=0.1$, island formation
with the condensed structure and a $(4 \times 4)$ periodicity
occurs as indicated by the dashed line.
This is explained in much the same way as was done
in the discussion of the condensation-into-islands phenomenon in
Section IV\,A\,1 for K/Al\,(111).

The theoretical  results  corresponding to
activated adsorption are collected in Fig.~\ref{fig8}b  where
for Na in the substitutional geometry
the adsorption, binding,  and vacancy formation energies are displayed.
It can be seen that
the adsorption energy for the substitutional geometry is
the most favorable for
 {\em all} coverages investigated. In particular,
substitutional adsorption with a
$(\sqrt{3} \times \sqrt{3})R30^{\circ}$
periodicity
has {\em the} most favorable adsorption energy.
It is therefore expected
that a condensation into $(\sqrt{3} \times \sqrt{3})R30^{\circ}$
islands with the  Na atoms in substitutional sites  occurs, beginning
at very low coverages, provided that thermal energy enables the transition
to occur. The atomic geometry is depicted in Fig.~9b; it is equivalent to
that of Fig.~7b with the only difference being the sizes of the adatoms.
The mechanism which stabilizes the
$(\sqrt{3} \times \sqrt{3})R30^{\circ}$
structure of
the substitutional adsorption is identified as
being the same as that for K on Al\,(111) \cite{neugebauer1}, i.e.,
as being due to the particular stability of the
$(\sqrt{3} \times \sqrt{3})R30^{\circ}$
surface vacancy structure.

Upon further deposition of Na
at room temperature onto the
$(\sqrt{3} \times \sqrt{3})R30^{\circ}$
substitutional adsorbate structure, the formation of a $(2 \times 2)$ surface
structure has
been observed experimentally.
This phase corresponds to
a coverage of $\Theta_{\rm Na}=1/2$ where there are
two Na atoms in the surface unit cell.
It has been the subject of
a number of studies, and many different models have been proposed.
Initially it was suggested
that the structure corresponds to three domains of
 $(2 \times 1)$ periodicity, involving Na atoms adsorbed on an unreconstructed
substrate \cite{porteus}.
 Subsequently it was argued that the structure
consists of two layers of Na atoms, each with
$(2 \times 2)$ periodicity, located on an
unreconstructed substrate
\cite{hohlfeld}.  Recent studies using HRCLS
\cite{andersen2},
normal incidence standing x-ray wave-field (NISXW)  and
surface extended x-ray adsorption fine structure (SEXAFS) \cite{kerkar2}
have surprisingly shown that a strong
intermixing of the Na and Al atoms in the surface
region occurs.
In the NISXW study
a model was proposed
involving two reconstructed layers each of stoichiometry NaAl$_{2}$
while from a recent scanning tunnel microscopy (STM)
 study \cite{brune1,brune2}
it was suggested that the $(2 \times 2)$ phase consists of
two Na layers where
one Na adatom  in the surface unit cell is located in
a surface substitutional site
and the second Na atom occupies an on-surface hollow site.
In order to clarify the situation regarding the
exact surface atomic
structure of this controversial phase, DFT-LDA calculations were
carried out for a large number of conceivable geometries \cite{stampfl4}.
These
calculations provided
evidence for the uniqueness of the structure and  permitted the rejection
of all of  the above-mentioned suggested models. The identified structure
is significantly more complex and unexpected than any of the
alkali metal adsorption systems studied so far. It
is drawn in Fig. \ref{fig9}c and
may be described as a composite double
layer ordered surface alloy and has the lowest total energy. In this structure
an Al atom is incorporated into a double Na layer
of which one of the Na atoms in the surface unit cell
occupies a substitutional
site and the second Na atom plus an Al atom are situated in
fcc- and hcp-hollow sites, respectively.
Alternatively,  it could be described as a four  layer Al-Na-Al-Na
alloy on an unreconstructed Al\,(111) surface.

Intuitively one might expect that the Al atom in the composite
double layer would adsorb in the fcc-hollow site rather than in the hcp-hollow
site (continuing growth of the bulk),
 and that the Na atom would assume the hcp-hollow site. It has
been shown from
the calculations that the reason this does not occur is due to
a larger relaxation of the reconstructed Al layer
(the Al layer containing three Al atoms per surface unit cell) and
of the substrate, compared to the structure where the atoms occupy the
opposite hollow sites. These
relaxations correspond to a significant
energy gain of 0.12~eV per surface unit cell.

The adsorption energy per Na atom in this surface alloy
structure is less than
that of the $(\sqrt{3} \times \sqrt{3})R30^{\circ}$
substitutional phase and greater than that of the non-activated
$(4 \times 4)$ surface geometry. The fact that it is
greater than the $(4 \times 4)$ structure explains why
the $(4 \times 4)$ structure is not found at room temperature.
The result that the $(2 \times 2)$ surface alloy
 structure has a smaller adsorption
energy than the
substitutional adsorbate $(\sqrt{3} \times \sqrt{3})R30^{\circ}$
structure, implies that
the $(2 \times 2)$ surface alloy will only form when it is not possible for the
$(\sqrt{3} \times \sqrt{3})R30^{\circ}$
structure to grow any further.

{}From consideration of the atomic structure of the
$(2 \times 2)$ phase, it would seem that no mass
transport is necessary in its formation.
However, the lower coverage
$(\sqrt{3} \times \sqrt{3})R30^{\circ}$
substitutional structure involves displacement of 1/3 of a monolayer of Al
atoms, which are assumed to diffuse across the surface
to be re-adsorbed at steps.
That is, the steps act as {\em sinks} for
the displaced substrate atoms. The results indicate,
 therefore, that formation of
the $(2 \times 2)$ structure from the
$(\sqrt{3} \times \sqrt{3})R30^{\circ}$
structure involves the reverse process, that is, diffusion of 1/3 of a
monolayer
of Al atoms back from the steps which are used in the
formation of
the $(2 \times 2)$ structure. In this case the steps serve  as
{\em sources} of substrate atoms \cite{burchhardt2}.

We note that the direct involvement of steps is one, but not
the only possibility. Another mechanism could be that the kicked out
Na atoms diffuse to regions between on-surface hollow-site Na
atoms. This mechanism gives rise simultaneously to two domains
of substitutional adsorption, one on the lower terrace from where the Al
atom was ejected and one on the upper terrace to where the ejected Al atom
finally resides. This two-domain mechanism has been discussed for
the Na adsorption on the Al\,(100) surface  \cite{berndt}.

It is appropriate to briefly  mention some of the recent experimental
studies.
The results of the  calculations for non-activated adsorption
are compatible with a SEXAFS study of
the $(4 \times 4)$ structure formed
at low temperature which found that the Na atoms  form a quasi-hexagonal
close-packed adlayer where, within the unit cell, there is a
slight clustering and rotation  of the Na adatoms with
respect to a strictly hexagonal close-packed arrangement \cite{schmalz2}.
In the calculations these small displacements of the Na atoms within
the $(4 \times 4)$ hexagonal adlayer were not considered.
For activated adsorption the results of the
DFT-LDA calculations are in full accordance with SEXAFS \cite{schmalz1},
LEED \cite{nielsen11}, and
NISXW \cite{kerkar1} studies of the
$(\sqrt{3} \times \sqrt{3})R30^{\circ}$
phase.
Nearest neighbor Na-Al distances in the substitutional geometry of
3.13, 3.21, 3.31, and 3.09~\AA\, were determined from the DFT-LDA
calculations,
LEED intensity,  SEXAFS, and NISXW studies, respectively.
The theoretical results are also
in accordance with STM pictures taken
at room temperature  in the coverage range below $\Theta_{\rm Na}=1/2$,
which showed that islands with a
$(\sqrt{3} \times \sqrt{3})R30^{\circ}$
structure are present at low coverages and increase in size with
coverage up to $\Theta_{\rm Na}=1/3$.
The ordering of the total energies of the $(\sqrt{3} \times
\sqrt{3})R30^{\circ}$ substitutional structure and the $(2 \times 2)$ four-
layer surface alloy structure
is consistent with an  STM study \cite{brune1,brune2} which shows that
initially
islands with a $(\sqrt{3} \times \sqrt{3})R30^{\circ}$
structure form rather than with a
$(2 \times 2)$ structure.

The complex structure of the four-layer surface alloy proposed in the DFT-LDA
study has been confirmed
by a detailed LEED intensity analysis \cite{burchhardt2}.
SEXAFS data are consistent with the
surface structure but the analysis does
not allow a unique discrimination against alternative
models  \cite{burchhardt2}.
In addition, results of an x-ray photoelectron diffraction (XPD) study
\cite{fasel2} also
agree with the predicted structure.
For the first three interlayer spacings the values determined by
DFT-LDA, LEED, SEXAFS, and XPD
agree to within 0.13~\AA\, \cite{comment1}.
The DFT-LDA and LEED results show, in particular,
remarkable agreement.
Both identify the presence
of small ($\approx$ 0.04~\AA\,) lateral displacements
of the Al atoms in layers four and five and small vertical displacements
of the Al atoms in layers
five and six (counting the uppermost Na layer as the first,
see Fig.~\ref{fig9}c).
The structure derived from the DFT-LDA total energy minimization is
at variance with that suggested by the STM analysis of Brune {\em et al.}
\cite{brune1,brune2} which is apparently due to the fact that STM is unable to
detect all surface atoms with equal clarity and to identify the
atomic species unambiguously.

\subsubsection{Atomic geometries of Na on Al\,(001) for $0 < \Theta_{\rm Na}
\protect\leq 0.5$}

The atomic density of the fcc (001) surface is
about 15\% lower than that of  the (111) surface,
and the (001) surface therefore has higher diffusion energy barriers to
adatoms.
As a consequence it is to be expected that the behavior of adsorbates
on the (001) surface may
exhibit some differences to on the (111) surface.
We will show and explain in this subsection that at
low temperatures Na adatoms always occupy on-surface
hollow sites with a repulsive intra layer interaction
and no indication of a phase transition (at low coverage)
into close-packed islands.
At higher temperature we find that
Na adatoms occupy at low coverage also on-surface hollow sites, but
at $\Theta_{\rm Na} \approx 0.15$ a transition into islands with
a substitutional geometry and a $c(2 \times 2)$ periodicity is predicted.
In contrast to the other sections, we
will discuss at greater length the results of  recent
experiments because the experimental situation and published
conclusions are more complex and contradictory than those for the other
systems.

DFT-LDA  calculations were carried out for many possible surface geometries
and coverages \cite{stampfl2,stampfl3} and some of these results are shown
in Fig.~\ref{fig11}a.

We start with a discussion of the on-surface geometries which are relevant for
non-activated adsorption. The  energies of different on-surface sites
differ more widely than those for Na on Al\,(111) which
can be attributed to the more corrugated potential energy surface of Al\,(001).
For the $c(2 \times 2)$ structure ($\Theta_{\rm Na}=0.5$)
the adsorption energies
for the on-top, bridge, and on-surface hollow
sites are 1.33, 1.45, and 1.59~eV \cite{stampfl2}. Therefore,
only the on-surface hollow sites were considered further, the other sites were
regarded as unlikely in the considered coverage range.
Figure~\ref{fig11}a shows that the on-surface hollow site
is clearly preferred over the substitutional site  at low coverages. Its
adsorption energy rapidly decreases
with increasing coverage, similar to K and Na on Al\,(111),
indicating a
strong repulsive interaction between the Na atoms.
However, differently to the adsorption on the (111) surface,
the theoretical results for Na on Al\,(001) give no indication
that a condensation into close-packed islands takes
place. This is quite certain for low coverage, but for
$\Theta  >  0.3$ we would not rule out that
a phase transition into $\Theta=0.5$ condensed islands can
occur. Figure~\ref{fig11}a  suggests
that the adlayer is compressed gradually up to
a coverage of $\Theta_{\rm Na} = 0.5$ for which Fig.~\ref{fig10}a
displays the structure. The difference  to the Al\,(111) surface arises
because the adsorbate-substrate bonding is stronger on Al\,(001) due to the
better embedding of the Na atom when occupying the
more open on-surface hollow site. It therefore remains energetically
advantageous
to keep those strong bonds with the substrate rather than to  build up
metallic bonds in a condensed adlayer island.

Figure~\ref{fig11}a also contains the results relevant for
activated adsorption, namely the adsorption
 energies of substitutional geometries.
It can be seen that the adsorption energy
for  Na in the surface substitutional site
depends much more weakly on coverage than
that for the on-surface Na, and in fact,
the adsorption energy for the $c(2 \times 2)$ substitutional structure,
which is displayed in Fig.~\ref{fig10}b,
is more
favorable than that of the lower coverage substitutional structures.
These results are interpreted as follows:
At low coverages
the on-surface adsorption in a homogeneous adlayer is the stable structure for
low as well as high temperature, but for higher
coverages on-surface adsorption becomes metastable.
For high temperature  adsorption
the adatoms then switch to substitutional sites,
forming islands with a $c(2 \times 2)$ structure (see Fig.~\ref{fig10}b)
which increase in size with increasing coverage.
The phase transition from the on-surface hollow site
to substitutional adsorption is indicated in Fig.~\ref{fig11}a
by the dashed line.

The shape of the adsorption energy versus coverage curve for
Na in the substitutional site ($E_{\rm ad}^{\rm sub}$ in Fig.~\ref{fig11}b)
appears, at first glance, somewhat similar
to the result for K and Na on Al\,(111) (see Figs.~\ref{fig6}b and
 \ref{fig8}b).
However, from a plot of the adsorption,
binding, and vacancy formation energies, which are also shown
Fig.~\ref{fig11}b, it can  be seen
that even though the {\em adsorption energy}  versus
coverage curve exhibits the
same trend as for K and Na on Al\,(111),
that is, in both cases
the most favorable adsorption energy is obtained
for the highest coverage structure,
the constituent components,
namely, the binding and vacancy formation energies
comply to a different behavior.
We see that on the (001) surface
the binding energy is particularly favorable for the
$c(2 \times 2)$ structure, reflecting an {\em attractive}
adsorbate-adsorbate interaction which actuates
island formation.
It can be seen in addition that
the surface vacancy formation energy shows {\em little}
dependence on surface vacancy concentration.

An additional surface structure was
considered
in the DFT-LDA calculations,
corresponding to coverage $\Theta= 0.5$ which
contains
Na atoms in on-surface hollow and in substitutional sites \cite{stampfl3}.
The surface unit
cell used was
twice that of the $c(2 \times 2)$ cell, i.e.,
 $(\sqrt{2} \times 2\sqrt{2})R45^{\circ}$.
In this arrangement the Na atoms form a quasi-hexagonal structure,
which
could also be described as alternating strips of Na atoms in
on-surface hollow and substitutional sites.
It was found that the  adsorption energy of this structure is between
that of the on-surface hollow and substitutional sites.
Accordingly,
similar mixed hollow-substitutional geometries
may well be present in the temperature and coverage ranges at which the
phase transition takes place, and they may also exist as
metastable surface geometries.

We now turn to a brief discussion of
experimental studies.
Early LEED intensity analyses \cite{hutchins,vanhove}
of $c(2 \times 2)$  Na/Al\,(001) concluded that the adatoms
occupy the on-surface hollow site. In the
preparation of the surface
from which the intensity versus energy curves were measured
 the sample was annealed at 360~K.
Recent experimental results are, however,
at variance with this conclusion. HRCLS \cite{andersen3} and SEXAFS
\cite{aminpirooz} studies demonstrated convincingly that the atomic structure
of the  $c(2 \times 2)$
phases are different when formed at low temperature and at room
temperature, that the structural phase transition is irreversible, and
that the room temperature structure is due to a mixing of Na and Al atoms.
The SEXAFS study determined
the
on-surface hollow site as the adsorption site at low temperature and
for the structure formed at room temperature
three possible models were proposed
with one being favored which consists of Na atoms in
on-top sites beneath a $c(2 \times 2)$ Al layer where the Al atoms
are adsorbed in
hollow sites with respect to the Na adlayer.
The DFT-LDA analysis of these three surface alloy structures
showed that they are energetically very
unfavorable, and these structures have therefore been
indubitably rejected \cite{stampfl3}.
Thus, these experimental results for the low temperature adsorption
agree with the
DFT-LDA results of Fig.~\ref{fig10}a. Also the experimental evidence
for a phase transition towards an intermixed structure agrees with the
theoretical prediction. However, the geometry which was derived from the
SEXAFS analysis is at clear variance with the DFT-LDA result shown in
Fig.~\ref{fig10}b.
Interestingly, in
an XPD study \cite{fasel1,fasel2} of the
room temperature $c(2 \times 2)$ phase of Na on Al\,(001)
it was concluded that the surface may contain
two domains, one with
Na atoms in the substitutional site (45\% occupancy)
and the other with the Na atoms
still in the on-surface hollow sites (55\% occupancy).

A recent dynamical LEED intensity analysis \cite{berndt}  shows that
the low temperature $c(2 \times 2)$ phase contains
Na in on-surface hollow sites and in the room temperature $c(2 \times 2)$ phase
the Na atoms adsorb substitutionally, in very good agreement with
the DFT-LDA  calculations.
Importantly, the LEED study reveals that
the transition from
on-surface hollow sites to substitutional sites proceeds in a gradual
step-by-step
way, beginning at a temperature of approximately 180~K,
and that on completion of the phase transition (determined to occur
at about 260~K)
the contribution to the LEED
intensities from substitutional adsorption is not less than 90\%.

As discussed
in Ref.~\cite{berndt},  one particular
atomic configuration consistent with the behavior of
the temperature dependent
satellite features, which are seen during the phase transition,
is one which corresponds to the creation of domains arranged
in  narrow
rows or ``strips'' where the Na atoms are in substitutional sites.
As the temperature is
increased and the phase transition proceeds, the rows or strips
of Na atoms in substitutional sites become progressively  wider until at the
end
of the transition all adatoms are in substitutional sites, recovering
again the  sharp $c(2 \times 2)$ LEED pattern \cite{berndt}.
 Three possible mechanisms
are envisaged
by which the substitutional geometry could be realized: $i)$
The Na atoms cause the ejection
of substrate Al atoms which diffuse across the surface to steps --
the steps acting as sinks. $ii)$ Steps on the surface act as sources
of Al atoms which diffuse {\em to} the regions between the on-surface
 hollow-site Na atoms,
forming exactly the substitutional geometry. $iii)$
As in $i)$ where the Na atoms eject substrate Al atoms which, however,
do not move to steps but diffuse as in $ii)$ to the regions between
on-surface hollow-site Na atoms. The latter mechanism gives rise simultaneously
to two regions of substitutional
adsorption, one on the ``lower'' terrace from where the Al atom was
ejected and one on an ``upper'' terrace to where the ejected Al atom finally
resides. ``Kick-out'' of the Al atoms by the Na atoms  could
proceed in a ``chain-reaction-like'' process which would give rise to
the formation of rows
or strips of Na atoms in substitutional sites, either on one terrace only
if process $i)$  occurs or on two terraces if  process $iii)$ occurs.
The starting point for the ``kick-out'' process could conceivably occur
at imperfections or steps on the surface.

The interatomic distances obtained by the various experiments
are in good agreement with those obtained from the DFT-LDA calculations.
For the on-surface hollow site,
the obtained values read 3.08, 3.21, 2.93, 3.27~\AA\, as obtained
by the DFT-LDA,  SEXAFS, XPD, and  LEED  studies.
For the room temperature phase
the Na-Al bond lengths determined
by the DFT-LDA, XPD, and  LEED investigations are
3.02, 3.11, and 3.07~\AA\, respectively.
For the substitutional site,
DFT-LDA and LEED identify a small
($\approx$ 0.05~\AA\,)
vertical displacement of the Al atoms in the second Al layer.
Both studies also find that
the effective radius of Na in the substitutional site is shorter than
in the on-surface
hollow site: approximately 4\% from DFT-LDA and 11\% from LEED.


In a manner similar to that described above for K on Al,
the temperature for the on-surface hollow--to--surface
substitutional transition has been estimated from DFT-LDA calculations.
The value obtained was 180~K and calculated
by assuming that the
transition state of the
reaction path is such that the Na atoms
are in substitutional sites and the ejected Al atoms are
in the energetically least favorable position  of the exchange diffusion
mechanism (two Al atoms above a surface vacancy,
see Refs.~\cite{stumpfPRB,feibelman}), not yet rebounded at kink
sites at steps.
This rough theoretical estimate of the transition temperature
is in good agreement with the value of 160~K determined from the HRCLS
study  \cite{andersen3} and with  180~K as determined in the
LEED study \cite{berndt}.

\subsection{Adsorbate bonding and electronic properties}

The widely differing adsorbate geometries and their strong
 dependencies on coverage
and temperature, as described in the previous section,
go together with  significant changes of the
surface chemical bond and surface electronic structure.
In this respect, in this section, we discuss the adsorbate-substrate and
adsorbate-adsorbate interaction
of the different surface structures, the associated electron densities at the
surface, and the resulting work function changes.

As mentioned in the introduction,
the characteristic change in
the work function with coverage, $\Delta \Phi (\Theta)$,
is typically explained in terms of the Gurney picture and
the ``usual form'' (see Fig.~\ref{fig2}) can be described reasonably
well within a model in which a high $r_s$ jellium
is adsorbed on $r_s=2$ jellium.
Alternatively, but with essentially the same conclusions, Muscat and Newns
\cite{muscat1,muscat3}
examined the coverage dependence of the
adsorbate-substrate and adsorbate-adsorbate interaction using an empirical
Anderson-type Hamiltonian \cite{anderson,newns}, which
takes the alkali metal valence states and the eigenstates
of the semi-infinite substrate  properly into account.
They studied the work function change for alkali metal
atoms on transition metal surfaces \cite{muscat3}.

To obtain
{\em quantitative}  agreement between calculated and experimental results
it is necessary to
take into account in the calculations that the atomic geometry (of adsorbate
and substrate)
and the surface electronic structure are closely linked.
Thus, a realistic description of the surface electronic properties
necessarily  requires a correct
treatment of the various  adsorbate arrangements discussed in Section IV\,A
above.
To highlight this point we investigate
the effect on the calculated adsorbate induced change of the work function
by assuming hypothetical atomic arrangements
of Na on Al\,(111) where the Na adatoms are homogeneously distributed over
the surface for the specified coverage.
The self-consistently calculated
surface dipole moments and work function change corresponding to these
structures
 are shown in Figs.~\ref{fig15}a  and \ref{fig15}b, respectively.
In agreement with the traditional picture of alkali metal adsorption,
there is a significant decrease of
the surface dipole moment with increasing
coverage.
For comparison, the results
for homogeneous adlayers of Na in the substitutional
site are also shown.
Here we see that the depolarization
effect, which is one indicator of the strength of the adatom-adatom
interaction,
is much less dramatic.
This is due to the fact that in the substitutional site the Na atoms sit
lower in the surface and the  repulsive
adsorbate-adsorbate interaction is screened better.
As a consequence of the lower (closer to the substrate) position
of the substitutional adsorbate we also see that
the dipole moment (at low coverage) is much smaller -- largely due to
the smaller dipole length.
As a consequence of this better screening of the adsorbate-adsorbate
interaction for substitutional adsorption, i.e., the (near)
absence of depolarization,  the $\Delta \Phi(\Theta)$
curve for the homogeneous substitutionally adsorbed adlayer does not exhibit
a minimum.
The obtained work function change
in Fig. \ref{fig15}b agrees reasonably well, as expected, with that of Lang's
jellium on jellium calculation.
However, as we will discuss below, the curves for Fig.~\ref{fig2}
and \ref{fig15}b are in clear
disagreement with the measured results.

\subsubsection{K on Al\,(111)}

The calculated change in work function of potassium on Al\,(111) corresponding
to the
realistic  atomic arrangements, which were discussed in Section IV\,A\,1,  are
displayed in Figs.~\ref{fig16}a and \ref{fig16}b for
non-activated and activated adsorption,
respectively. For the non-activated adsorption the adatoms stay on the surface,
but at $\Theta \approx 0.15$ they change from a homogeneous adlayer
to close-packed islands
(the structure was shown in Fig.~\ref{fig7}a),
which for increasing coverage increase in size. This phase transition is the
reason for the
minimum in the dashed curve of Fig.~\ref{fig16}a. The comparison with the
low temperature experimental result shows reasonable agreement.
In particular we emphasize that the origin of the minimum
in $\Delta \Phi(\Theta)$
is not due to the adsorbate-adsorbate depolarization, but to the
phase transition into close-packed islands.
It is interesting to inspect the adsorbate-induced change of the electron
density (see  Fig.~\ref{fig12}a).
Within this structure, the adsorbate-adsorbate distance is only 5\% larger
than that of bulk K, and  Fig.~\ref{fig12}a indeed reveals that
there is considerable charge density between the K adatoms. We therefore tend
to classify the bonding within the adlayer as metallic.
Consistent with this we note that the surface band structure of this
adlayer is similar to that of  a free-standing $(\sqrt{3} \times
\sqrt{3})R30^{\circ}$-K layer \cite{wenzien}.

The calculated work function change for activated adsorption
is shown in Fig.~\ref{fig16}b. Although this curve looks similar to that of the
low temperature (non-activated) adsorption, this similarity is in fact
highly incidental. The physics behind the curve of Fig.~\ref{fig16}b is
that of a structural phase transition at $\Theta \approx 0.15$ into
islands with substitutional adatoms (compare Fig.~\ref{fig7}b).
Incidentally, this phase transition occurs
at about the same coverage as the phase transition of the non-activated
adsorption which implies that the minima of the
$\Delta \Phi(\Theta)$ curves of Figs.~\ref{fig16}a and \ref{fig16}b
are at about the
same positions. The  charge density between the adsorbates (see
Fig.~\ref{fig12}b),
which is essentially provided by the substrate atoms,
is significant and screens the adsorbate-adsorbate interaction
quite efficiently.
Therefore the substitutionally adsorbed potassium can attain
its ionic character even in the close-packed
$(\sqrt{3} \times \sqrt{3})R30^{\circ}$ structure in clear contrast to
the on-surface adsorbate structures.
The DOS for the substitutional adsorbate (not shown) is also significantly
different to that of the on-surface adsorbate structure,
and it is mainly due
to the reconstruction of the Al substrate, i.e., to the removal of one third
of all surface atoms; the K atoms which occupy the surface vacancies
modify the vacancy electronic structure only little  \cite{wenzien}.

\subsubsection{Na on Al\,(111)}

The behavior of Na on Al\,(111) is qualitatively similar to that of
K on Al\,(111). However, because of the smaller size of the Na atom,
the condensed structure of the non-activated adsorption is more dense
for Na than for K. The
$(4 \times 4)$ structure of the Na islands (local coverage $\Theta = 9/16$),
which is assumed for low temperature adsorption and coverages above $\Theta
\approx 0.1$ was displayed in Fig.~\ref{fig9}a.
The calculated work function change is
shown in Fig.~\ref{fig17} for different temperatures. The minimum at $\Theta
\approx 0.1$ marks the
phase transition from the homogeneous adlayer to the condensed
islands which grow in size
with increasing coverage. Again we emphasize that the shape of the
$\Delta \Phi(\Theta)$ curve and the physics behind this shape is
significantly different to that of the ``traditional'' picture of
Figs.~\ref{fig2} and  \ref{fig15}b.
In the condensed
structure the nearest-neighbor Na-Na distance is very similar to
that in bulk Na and
the intra layer bonding can be described as metallic.

The calculated work function change with coverage for the activated adsorption
follows a linear relation between $\Delta \Phi$ and $\Theta$
(see Fig.~\ref{fig17}b).
This results, because at high temperature island formation starts at
very low coverage and thus the work function change is mainly determined by
the surface
dipole moment of the condensed
$(\sqrt{3} \times \sqrt{3})R30^{\circ}$-Na substitutional structure
times the number of Na atoms in these islands. The work function decreases
up to the point where the whole surface is covered with $(\sqrt{3} \times
\sqrt{3})R30^{\circ}$  substitutional Na atoms.
For coverages higher than $\Theta = 1/3$ (compare Section IV\,A\,2) a major
restructuring of the
surface takes place. The
analysis of the work function shows
that $\Delta \Phi (\Theta)$ should exhibit a minimum at $\Theta = 1/3$
($\Delta \Phi  \approx -1.6$~eV) and then
increase up to $\Delta \Phi(\Theta)= -1.16$ eV at $\Theta = 1/2$ for
the four-layer surface alloy.
For even higher coverages
the curve should reach the value of
the work function of a Na crystal surface.

Similar to that of K in the substitutional site, the bonding of
Na in the $(\sqrt{3} \times \sqrt{3})R30^{\circ}$ substitutional structure
can be described as partially ionic.
The density difference plot is shown in Fig.~\ref{fig122}.

On further deposition of one sixth of a monolayer of Na onto the substitutional
$(\sqrt{3} \times \sqrt{3})R30^{\circ}$-Na surface, the
$(2 \times 2)$ surface alloy structure forms.
The adsorbate induced electron density is shown in Fig.~\ref{fig133}.
Two regions of charge density maxima occur. Both are weighted towards
the uppermost  Al atom. Inspection of the
{\em total} charge density reveals that the Al atom
in the hcp-hollow site forms covalent-like
bonds with the three Al atoms under it.
This is possible because the Al atoms in the reconstructed layer underneath
are rather lowly coordinated due to the displacement of one
quarter of a monolayer of Al atoms.
The Al-Al bondlength is shorter than in bulk Al, compare 2.66 to
2.81~\AA\ (theoretical bulk value) reflecting the stronger
and lower coordinated bonding.
The {\em occupied} adsorbate induced  DOS for the $(2 \times 2)$ structure
is largely determined by the Al structure, which corresponds
to
the Al\,(111) surface with $\Theta = 1/4$ vacancies plus
$\Theta = 1/4$ Al atoms in the on-surface hcp-hollow
sites (compare Fig.~\ref{fig9}c).
In particular we noted a  peak at approximately 2~eV below the Fermi level
(see Ref.~\cite{stampfl4}).

\subsubsection{Na on Al\,(001)}

Considering now Na on Al\,(001), we see that from
inspection of the {\em total} charge density for the $c(2 \times 2)$
structure with Na adsorbed substitutionally, that
directional bonds are formed between  Al atoms of the uppermost
 layer (from which
$\Theta=0.5$ Al atoms have been removed)
and their  four  nearest-neighbor
atoms in the layer underneath.
The density difference plot in Fig.~\ref{fig13}
clearly displays a maximum of charge density
located above the uppermost Al atoms.
{}From inspection of the adsorbate induced
DOS,  an occupied and an
unoccupied feature at about 2~eV below and 1~eV above the Fermi level,
 respectively are identified. While the  peak below the Fermi level is
largely due to the Al structure (substrate with vacancies),
the peak above the Fermi level arises from the Na atoms.
An analysis of the wave functions of these states
indicated that (at least at $\overline{\Gamma}$)  the occupied state is
localized above the
uppermost Al atoms and the unoccupied state is localized above the
Na atoms.


\section{Conclusion}

In the present paper we discussed
results of recent density functional theory calculations for alkali metal
adsorbates on close-packed metal surfaces.
The analyses support
the traditional charge transfer picture of
alkali metal adsorption as being
an appropriate description  in the limit of low coverage
and non-activated adsorption, i.e., when isolated alkali metal adatoms occupy
on-surface positions.
It is emphasized, however, that
low coverage means {\em very} low
because adsorbate-adsorbate interactions can become very noticeable
already at surprisingly low coverages
(e.g. $\Theta < 0.1$)
and the traditional picture, thus,  should not
be generally applied in the higher
coverage regime. In particular we stressed the interplay of the electronic
properties of the surface, which is for example reflected in the work function,
and the adsorbate site.

The transition region from the regime at very low coverages,
governed by electrostatic repulsion, to
higher coverage regimes where adlayer phases develop
which possess a cohesive   interaction  between the adatoms
(or vacancies) is particularly interesting.
In this respect, it has been shown
that with increasing coverage,
Na and K adsorbates  on close-packed Al surfaces
exhibit an unexpected variety
of physical and chemical phenomena. For non-activated adsorption
these phenomena
include site changes and on-surface condensation
phase transitions and for activated adsorption,
condensation phase transitions
involving substitutional adsorption, as well as surface alloy
formation.
Such unusual and varied  behavior as described in the present work
with respect to the knowledge of several years ago,
emphasizes a new updated view  of alkali
metal adsorption on metal surfaces.

The novel adsorption geometries were observed so far for the two
close packed aluminum surfaces: Al\,(111) and Al\,(100).
Thus, the metal, which so far was assumed to be the most
jellium-like system, reveals in fact quite clearly  the influence of
its atomic structure and tendency towards
geometries which are stabilized by the formation of
partially covalent bonds. At this point it is not clear,
whether similar behavior ought to be expected for other
systems. Calculations of the formation energy of surface vacancies \cite{pola}
indicate that silver and maybe also copper
are further candidates for substrates where substitutional
adsorption might occur. In fact, substitutional adsorption
is a much more general phenomenon than hitherto expected.
Atoms which are immiscible with the bulk may well intermix at the
surface, because there the size mismatch can be easy accounted for
by relaxations.


\begin{figure}
\caption{Schematic representation of the emptying and
broadening of the alkali metal atomic $s$ level
as it approaches a metal surface.
\protect$\Phi_{s}$ and  \protect$\Delta \Phi$ represent
the substrate work function and the change in the substrate work function,
respectively, and \protect$ E_{F}$  and \protect${I}_{a}$
denote the Fermi level of the substrate and the first ionization
energy of the free adatom. The figure is drawn with values which resemble
the adsorption of Na on Al\,(111) at low temperature and low coverage
($\protect\Theta \protect\approx 0.1$).}
\label{fig1}
\end{figure}

\begin{figure}
\caption{Change in work function with increasing coverage
obtained by Lang \protect\cite{lang1} using a jellium on jellium model
for parameters corresponding to Na on Al\,(111).
}
\label{fig2}
\end{figure}

\begin{figure}
\caption{
(a) Change in the density of states due to adsorption of sodium (dashed),
silicon (dot-dashed), and chlorine (dotted) on Al\,(111)
 (from Ref.~\protect\cite{bormet1}).
The density of states of the aluminum substrate is shown as a solid line.
Energies are given with respect to the Fermi level which is indicated
by a vertical full line. The vertical dashed line indicates the bottom of
the valence band.
(b) Analogous quantities to (a)
for a jellium substrate with an electron density which is approximately equal
to the average value of Al: $r_{s}=2$ bohr.
 Li is considered instead of Na
(after Fig.~2 of Ref.~\protect\cite{lang3}).
}
\label{fig3}
\end{figure}

\begin{figure}
\caption{
Total valence electron density in a plane perpendicular to the
surface in the $[1 \overline{2} 1]$ direction for the
chemisorption of sodium, silicon, and chlorine
(from left to right) on Al\,(111).
The contour spacing is $ 29 \times (1.5)^{k} \times 10^{-3}
{\rm bohr}^{-3}$ with $-5 \leq k \leq +5$. The green-yellow
boarder line $(k=0$, $n({\bf r}) = 29 \times 10^{-3}
{\rm bohr}^{-3}$ is the average density of the aluminum, i.e., $r_{s}=$
2.02 bohr.
}
\label{fig4}
\end{figure}

\begin{figure}
\caption{
Density difference, $n({\bf r}) - n^{0}({\bf r}) - n^{\rm atom}({\bf r})$,
for Na, Si, and Cl on Cu\,(111)
(upper panel) and on Al\,(111) (lower panel)
in a plane perpendicular to the
surface in the $[1 \overline{2} 1]$ direction (from Ref.~\protect\cite{yang}).
$n({\bf r})$, $n^{0}({\bf r})$, and $n^{\rm atom}({\bf r})$ are the
electron densities of the adsorbate system, the clean surface,
and the free, neutral adatom.
The hatched areas correspond to regions of charge increase and the unhatched
areas correspond to charge depletion. The units
are  10$^{-3}$ bohr$^{-3}$ and the contour lines  correspond to
the values  $-$5.0, $-$2.5, $-$0.5, 0.0, 0.5, 2.5, and 5.0 .
}
\label{fig5}
\end{figure}

\begin{figure}
\caption{
(a) Adsorption energy versus coverage for K in on-top and
fcc-hollow sites on Al\,(111). The dashed line marks the phase
transition from the homogeneous adlayer into adatom islands with a
$(\protect\sqrt{3} \protect\times \protect\sqrt{3}) R 30^{\protect\circ}$
periodicity (from  Ref.
\protect\cite{neugebauer1}).  (b) Adsorption,
$E_{\rm ad}^{\rm sub}$, and binding energies, $E_{b}^{\rm sub}$,
for substitutional adsorption, and vacancy formation energy,
$E_{f}^{\rm vac}$, versus coverage (from Ref.
\protect\cite{neugebauer3}
).
}
\label{fig6}
\end{figure}

\begin{figure}
\caption{
Perspective view of the atomic structure of
$(\protect\sqrt{3} \protect\times \protect\sqrt{3}) R 30^{\protect\circ}$-K
on Al\,(111).
(a) K atoms (red) are in on-top sites and the substrate
atoms under the potassium atoms (yellow) are displaced towards the
bulk and the substrate atoms between the potassium
atoms (blue) are displaced upwards with respect
to the ideal positions.
(b) Substitutional geometry where the
Al atoms which were beneath the K atoms in (a) have been kicked
out.
}
\label{fig7}
\end{figure}

\begin{figure}
\caption{
Adsorption energy versus coverage for Na in on-top and fcc-hollow
sites on Al\,(111). The dashed line marks the phase transition from the
homogeneous adlayer into adatom
islands with a condensed structure  (from Ref.~\protect\cite{neugebauer1}).
  (b) Adsorption, \protect$E_{\rm ad}^{\rm sub}$,
and binding energies, $E_{b}^{\rm sub}$,
 for substitutional adsorption, and
vacancy formation energy, $E_{f}^{\rm vac}$,
 versus coverage (from Ref.~\protect\cite{neugebauer3}).
}
\label{fig8}
\end{figure}


\begin{figure}
\caption{
Perspective view of (a) the condensed hexagonal
$(4 \protect\times 4)$-Na/Al\,(111)
surface structure, (b)
$(\protect\sqrt{3} \protect\times \protect\sqrt{3}) R 30^{\protect\circ}$-
Na/Al\,(111)
with Na in the surface substitutional site, and
(c) the $(2 \protect\times 2)$-Na/Al\,(111) surface alloy structure. The
purple, blue, and green circles represent, respectively,
            Na atoms and Al atoms in the first and second layer. In (c)
the Al atom in the hcp-hollow site of the surface alloy is shown in yellow
and the inequivalent Na atoms are depicted by different shades of purple.}
\label{fig9}
\end{figure}

\begin{figure}
\caption{ (a) Adsorption energy versus coverage
              for Na on Al\,(001) in the on-surface hollow site
              and in the
              surface substitutional site.
           (b) Adsorption, \protect$E_{\rm ad}^{\rm sub}$,
               and binding energies, \protect$E_{b}^{\rm sub}$,
               for
               substitutional adsorption, and
               the vacancy formation energy, \protect$E_{f}^{\rm vac}$,
                versus
               coverage (from Ref.~\protect\cite{stampfl3}).}
\label{fig11}
\end{figure}

\begin{figure}
\caption{
Perspective view of the atomic structure of
 $c(2 \protect\times 2)$-Na on Al\,(001).
The purple, blue, and green circles represent, respectively,
            Na atoms and Al atoms in the first and second layer.
(a) On-surface hollow site geometry and
(b) substitutional geometry, where every second
Al atom in the top unreconstructed Al layer has been kicked out.
}
\label{fig10}
\end{figure}

\begin{figure}
\caption{(a) Surface dipole moment for homogeneous adlayers
of Na on Al\,(111) in the fcc-hollow site and
in the surface substitutional site
versus coverage.
(b) The corresponding change in the
work function where the
same symbols as in (a) are used (after Ref.~\protect\cite{neugebauer3}).}
\label{fig15}
\end{figure}

\begin{figure}
\caption{Calculated (dashed line) and measured (full line with open
circles) work function change, $\Delta \Phi$,
with coverage for K on Al\,(111). (a)
For ``low temperature''
and (b) for ``high temperature''
(from Ref.~\protect\cite{neugebauer1}).
The experimental results are from Ref.~\protect\cite{horn}.}
\label{fig16}
\end{figure}

\begin{figure}
\caption{ Change of the electron density for
$(\protect\sqrt{3} \protect\times \protect\sqrt{3})R30^{\protect\circ}$-K
 on Al\,(111) for K  in (a) the on-top site and
(b) the substitutional site.
For the on-top site the reference system
is the clean surface and for the substitutional site
it is the
 $(\protect\sqrt{3} \protect\times \protect\sqrt{3})R30^{\protect\circ}$
surface vacancy structure. The contours are in the plane normal to the
surface and parallel to the $[1 \overline{2} 1]$ direction. The substrate atoms
are represented by the small dots and the K atoms by the large dots.
The units are $10^{- 3}$ bohr$^{-3}$
(from Ref.~\protect\cite{neugebauer2}).}
\label{fig12}
\end{figure}

\begin{figure}
\caption{Calculated (dashed line) and measured (full line with open
circles) work function change, $\Delta \Phi$,
with coverage for Na on Al\,(111). (a)
For ``low temperature''
and (b) for ``high temperature''
(from Ref.~\protect\cite{neugebauer1}).
The experimental results are from Ref.~\protect\cite{hohlfeld}.}
\label{fig17}
\end{figure}

\begin{figure}
\caption{ Change of the electron density for
$(\protect\sqrt{3} \protect\times \protect\sqrt{3})R30^{\protect\circ}$-Na
 on Al\,(111) for Na in the substitutional site.
The reference system is the
 $(\protect\sqrt{3} \protect\times \protect\sqrt{3})R30^{\protect\circ}$
surface vacancy structure. The contours are in the plane normal to the
surface and parallel to the $[1 \overline{2} 1]$ direction. The substrate atoms
are represented by the small dots and the Na atoms by the large dots.
The units are $10^{- 3}$ bohr$^{-3}$
(from Ref.~\protect\cite{neugebauer2}).}
\label{fig122}
\end{figure}

\begin{figure}
\caption{Change of the electron density for the
$(2 \protect\times 2)$-Na/Al\,(111)
surface alloy structure
The reference system is the Al\,(111) surface
plus the reconstructed Al layer and the Al atom in the hcp-hollow site
on the reconstructed layer.
The contours are in the plane normal to the
surface and parallel to the $[\overline{1} \overline{1} 2]$
direction.  The Al atoms
are represented by the small dots and the Na atoms by the large dots.
The units are $10^{-3}$ bohr$^{-3}$
(from Refs.~\protect\cite{stampfl4}).}
\label{fig133}
\end{figure}

\begin{figure}
\caption{Change of the electron density for
$c(2 \protect\times 2)$-Na/Al\,(001)
with the Na atoms in the substitutional site.
The reference system is the
$c(2 \protect\times 2)$ surface vacancy structure.
The contours are in the plane normal to the
surface and parallel to the
[110] direction.  The Al atoms
are represented by the small dots and the Na atoms by the large dots.
The units are $10^{-3}$ bohr$^{-3}$
(from \protect\cite{stampfl3}).}
\label{fig13}
\end{figure}



\begin{references}
\bibitem{jellium} We note that the jellium model has also
some severe limitations, as for example a negative surface energy
for most metals (i.e., a density parameter $r_{s} \protect\lesssim$ 2.5 bohr).
\bibitem{sign} Note that in an earlier work
\cite{neugebauer2,neugebauer1} the adsorption and binding
energies (equs. \ref{eq3_1}, \ref{eq3_2}, \ref{eq3_4}) were defined with
opposite sign. Throughout this paper we adopt the sign
convention that for an exothermic process the adsorption energy
is positive.
\bibitem{comment1} In the XPD study the third interlayer spacing
 was assumed to be the same as the first interlayer spacing of the
$(\sqrt{3} \times \sqrt{3})R30^{\circ}$-Na/Al\,(111) structure.
\end{references}
\end{document}